\documentclass{article}
\usepackage{ijcai23}

\usepackage{times}
\usepackage{soul}
\usepackage{url}
\usepackage[hidelinks]{hyperref}
\usepackage[utf8]{inputenc}
\usepackage[small]{caption}
\usepackage{graphicx}
\usepackage{amsmath}
\usepackage{amsthm}
\usepackage{booktabs}
\usepackage{algorithm}
\usepackage{algorithmic}
\usepackage[switch]{lineno}

\title{Privacy-Preserving End-to-End Spoken Language Understanding}

\author{
Yinggui Wang$^1$$^*$
\and
Wei Huang$^1$ \thanks{These authors contributed equally to this work} \And
Le Yang$^{2}$
\affiliations
$^1$Ant Group\\
$^2$University of Canterbury
\emails
wyinggui@gmail.com, 
hw378176@antgroup.com, 
le.yang@canterbury.ac.nz
}

\begin{document}

\maketitle

\begin{abstract}
    Spoken language understanding (SLU), one of the key enabling technologies for human-computer interaction in IoT devices, provides an easy-to-use user interface. Human speech can contain a lot of user-sensitive information, such as gender, identity, and sensitive content. New types of security and privacy breaches have thus emerged. Users do not want to expose their personal sensitive information to malicious attacks by untrusted third parties. Thus, the SLU system needs to ensure that a potential malicious attacker cannot deduce the sensitive attributes of the users, while it should avoid greatly compromising the SLU accuracy. To address the above challenge, this paper proposes a novel SLU multi-task privacy-preserving model to prevent both the speech recognition (ASR) and identity recognition (IR) attacks. The model uses the hidden layer separation technique so that SLU information is distributed only in a specific portion of the hidden layer, and the other two types of information are removed to obtain a privacy-secure hidden layer. In order to achieve good balance between efficiency and privacy, we introduce a new mechanism of model pre-training, namely joint adversarial training, to further enhance the user privacy. Experiments over two SLU datasets show that the proposed method can reduce the accuracy of both the ASR and IR attacks close to that of a random guess, while leaving the SLU performance largely unaffected.
\end{abstract}

\section{Introduction}

\begin{figure*}
	\begin{center}
		\includegraphics[width=0.65\linewidth]{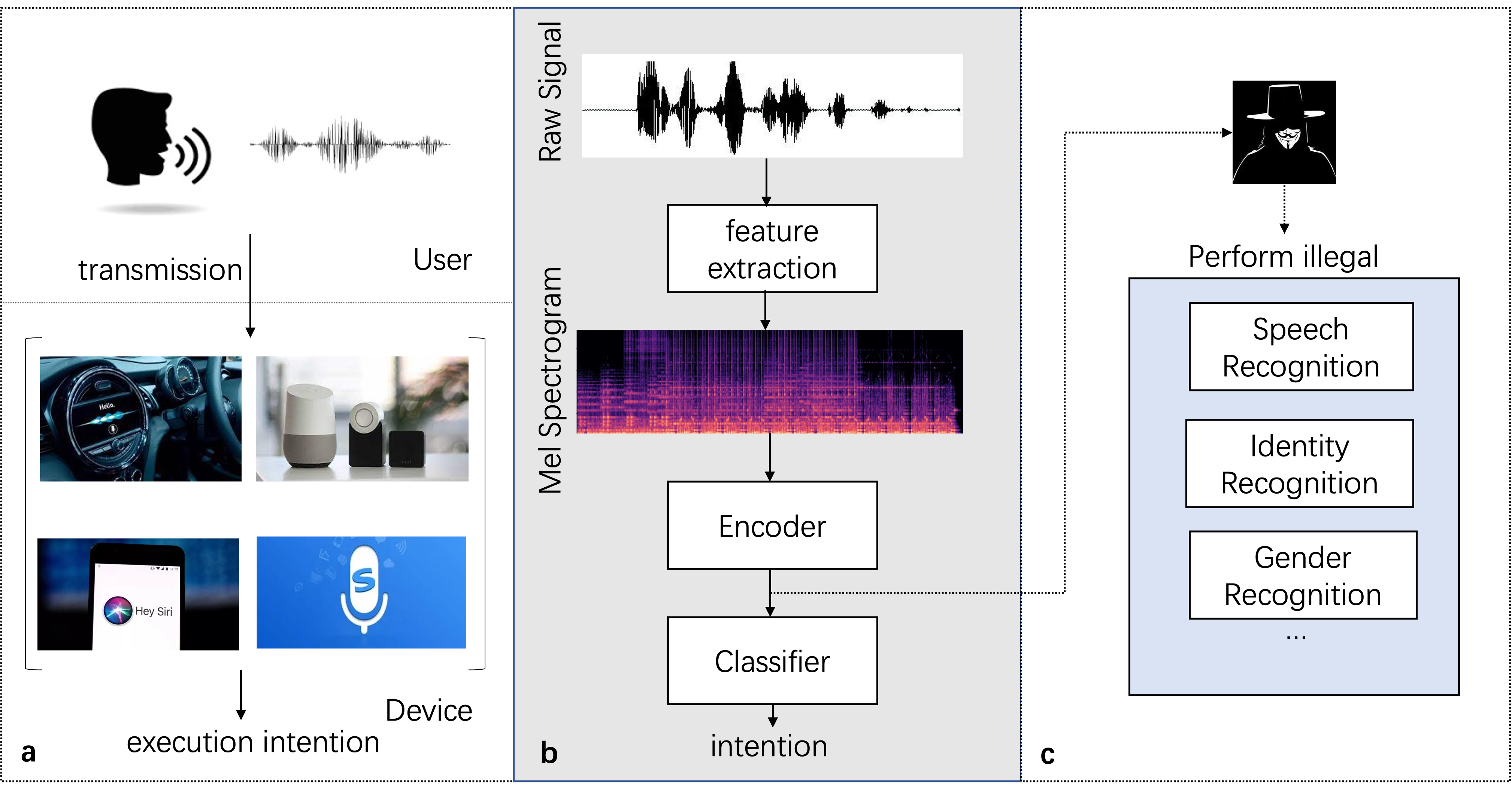}
	\end{center}
	\caption{(a) Speech Control System, (b) End-to-End SLU, (c) Potential Malicious Inference Attack.}
	\label{fig:picture1}
    \end{figure*}

    Voice-controlled IoT devices and smart home assistants are gaining popularity in people's lives, and spoken language understanding (SLU) has become one of the main enabling technologies to achieve human-machine interaction between users and devices \cite{coucke2018snips}. For example, SLU can be found in the in-car voice system of Tesla electricity vehicles (EVs), the voice assistant of Apple phones and Xiaomi's smart home ecology. Figure \ref{fig:picture1}(a) shows how these systems work in general.

	There are certain privacy risks associated with these IOT devices \cite{atlam2020iot}. Due to their limited storage space, some vendors choose to deploy only the encoder of the model at the client side and transmit the output of the encoder to the server side for performing tasks such as prediction and classification, which gives opportunities for malicious attackers to take advantage of. Recent end-to-end SLU systems predict user intent directly from speech \cite{serdyuk2018towards} (see also Figure \ref{fig:picture1} (b)). This approach circumvents the conversion of speech to text, and to a certain extent, it avoids privacy leakage caused by text theft. However, strictly speaking, end-to-end SLU systems are still subject to certain risks of privacy breach. As depicted in Figure \ref{fig:picture1}(c), the recognition model may be vulnerable to being stolen by an untrusted third party and used to deduce the sensitive information about the user. In other words, to maintain sensitive user information, the SLU systems need to be able to infer the user's intent from the masked voice only, while the specific voice content, identity information and other user-sensitive attributes should be dispensed.

    With the increasing public concern about the protection of personal privacy information and higher requirements of national standards on the protection of biological information, significant efforts have been invested to address the problem of user privacy leakage. The utilization of cryptography, multi-party secure computing, differential privacy, and federation learning has been widely investigated \cite{xu2021privacy}. However, these methods have their own drawbacks. For example, using cryptography or differential privacy has a great impact on the classification accuracy of the model. Besides, the voice data being more complex leads to less efficient computation for multi-party secure computation, while under the federated learning, the end-side computation capability is limited. With the increasing refinement of deep learning methods, various new schemes are developed with adversarial training and disentangled representation being the more popular choices.
    
    The current privacy-preserving intention understanding algorithms normally protect the voice print information of the speaker only, leaving the content information unprotected. From the perspective of privacy protection, this is not satisfactory. Intent understanding is usually used in the smart home and human-computer interaction scenarios, both of which use voice as the control signal. In those cases, wake-up words are sometimes used as a prompt for issuing instructions. When judging whether the speaker’s voice contains wake-up words, the system needs to record voice in real time, and then deduces the intent when it finds that there are wake-up words in the recorded voice. Nevertheless, the intent of the voice may also be recognized directly without using wake-up words. In this case, the recording will be longer than usual. If the speaker mentions something involving privacy during this time, the content of the voice would involve privacy information. Therefore, the privacy-preserving intention understanding algorithm needs to protect both the voice print and content information at the same time.
    \begin{figure}[t]
	\centering
	\includegraphics[width=0.8\linewidth]{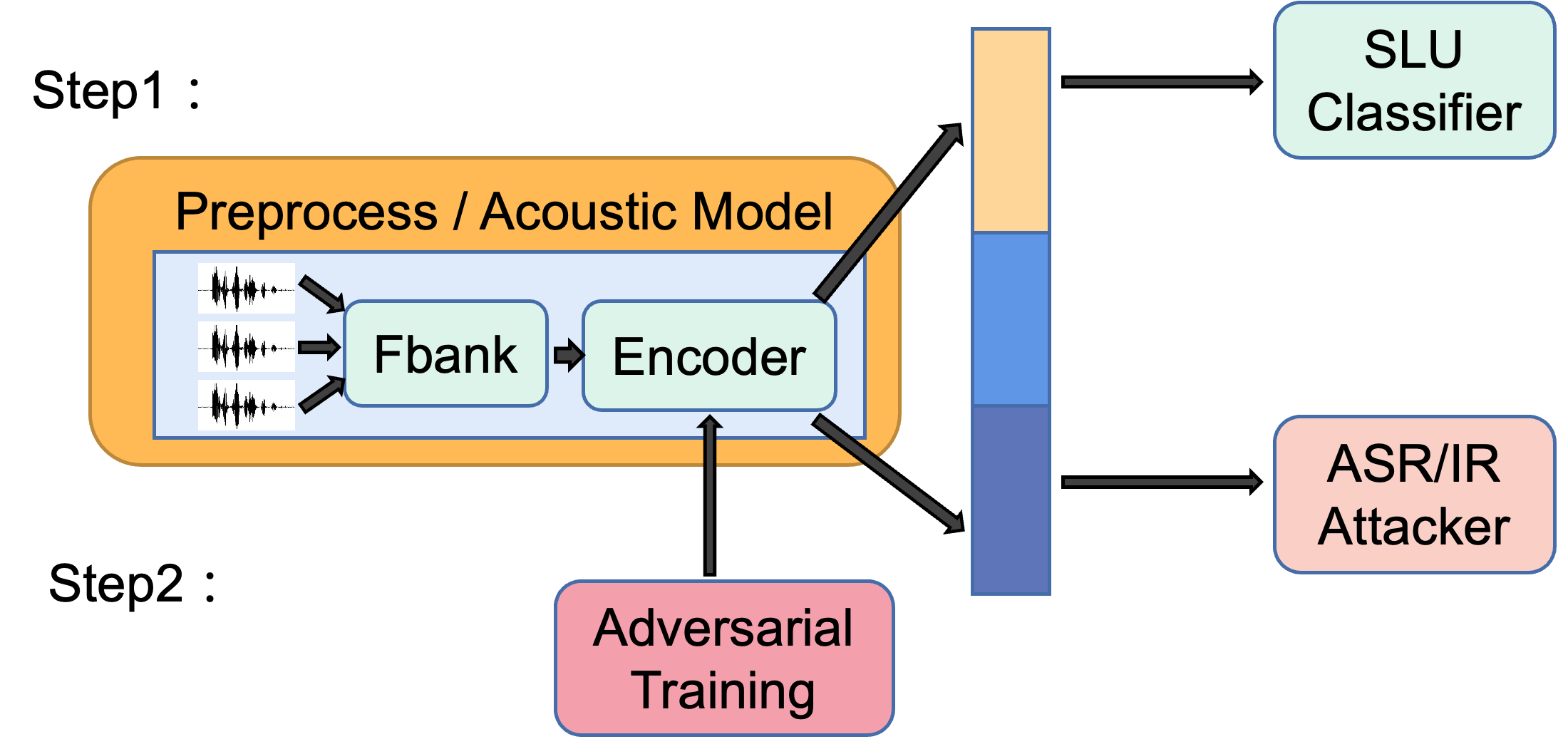}
	\caption{Framework of SLU privacy protection.}
	\label{fig:picture2}
    \end{figure}

    Inspired by the adversarial training \cite{timm2021privacy} and disentangled representations \cite{aloufi2020privacy}, this paper proposes a novel model integrating hidden layer separation and adversarial training. Figure \ref{fig:picture2} illustrates the developed privacy-preserving framework for SLU. First, we note that under the multi-task training, the information of the three tasks in consideration (i.e., SLU, automatic speech recognition (ASR) and identity recognition (IR)) in the encoder and hidden layers of the model would be entangled. This results in the attackers being potentially capable of inferring sensitive information for ASR as well as IR from the hidden layer output. We thus propose to partition the hidden layer of the model into several parts. The SLU task uses only one part of the hidden layer during training, while some other parts are reserved for ASR and IR processing. We further notice that there is some information similarity across the three considered tasks. The developed model is then augmented with the shared part in the hidden layer for the three tasks and the cosine similarity is utilized to impose isolation of the individual parts for the three tasks. With the above hidden layer separation, the information for SLU, ASR and IR are no longer mixed together. Instead, they occupy a separated part of the hidden layer. Specifically, only a specific part of the hidden layer contains useful information for SLU, but that part does not contain or contains only little information valuable for ASR and IR. This achieves the purpose of protecting privacy.

    To further enhance the privacy preservation ability of the proposed model, two approaches are employed in the inference of the model. We 1) use robust ASR and IR attackers to reason about sensitive information, and we 2) train ASR and IR attackers from scratch under the premise of freezing the privatizer (encoder) and observe whether sensitive information can be recovered.

    We verify the validity of the model framework using different datasets. Experimental results show that our framework not only guarantees the accuracy of the SLU task, but also reduces the success rate of attackers to close to that of a random guess. It also has better privacy protection than the popular adversarial training methods currently available.

    The main contributions of this paper are:
	\begin{itemize}
		 \item We demonstrate the potential vulnerability of deep learning-based SLU models, where a malicious attacker can accurately reason about sensitive user attributes from a  model without privacy protection.
		 \item To solve the above problem, this paper proposes a privacy-preserving framework that combines hidden layer separation with adversarial training. It partitions the hidden layer where multiple information is mixed together, so that the SLU task is carried only in a specific portion of the hidden layer, and redundant sensitive information is removed from that part for the purpose of privacy preservation.
		 \item By verifying the security of the proposed framework using both pre-trained robust attackers and re-trained test scenarios, experiments show that our framework balances efficiency and privacy protection well.
	\end{itemize}

\section{Related Work}
    \subsection{Spoken Language Understanding (SLU)}
    Early SLU tasks were carried out using two models. First, the speech sentences are input to the ASR model to obtain the text output, and then the text is input to the natural language unit (NLU) model for intent identification. However, this method is complex and difficult to tune. Thanks to the powerful abstraction capabilities of deep learning models, this solution is now gradually being replaced by the end-to-end SLU systems, which predict intent directly from the input speech fragments without going through generating intermediate text \cite{serdyuk2018towards,haghani2018audio}. In terms of model architecture, the end-to-end SLU framework has evolved from the original one that only uses classifiers to differentiate intent to one that uses ASR models for parameter initialization, complex models as feature extractors, and multi-task training with ASR to improve the classification accuracy \cite{arora2022espnet}.

    \subsection{Voice Privacy Protection}

    In this subsection, we briefly survey the techniques for voice privacy protection, which can be broadly categorized into cryptography-based methods and machine learning-based approaches without cryptography. Since the cryptography-based methods are not only time consuming but also complex in principle, machine learning-based approaches are more suitable for practical applications. With the development of deep learning, it has made a big splash in the field of voice privacy protection. 
    
    \cite{aloufi2020privacy} uses disentangled representation learning explicitly and extracts independent factors from the raw speech data, enabling different branches of the model to learn information for different tasks. It applies the variational auto-encoders (VAE) to reconstruct the raw data to ensure that sensitive user information is removed before sharing the speech data with cloud service providers. \cite{aloufi2020paralinguistic} later combines the model pruning, mixed-accuracy training, and quantization to further improve the utility of the model. \cite{stoidis2021protecting} also uses disentangled learning to encode information about different attributes into separate subspaces that can be decomposed independently, and applies the vector-quantised VAE (VQ-VAE) to reconstruct the data to protect gender as well as identity sensitive attributes. \cite{aloufi2021configurable} proposes the use of speech separation plus discretization to avoid the speaker information leakage problem due to overlapping speech. \cite{noe2020adversarial} verifies that the x-vector features extracted by speaker recognition systems are vulnerable to privacy leakage by untrusted third parties, and its then combines adversarial training and reconstruction algorithms to remove gender information from the original features for privacy preservation purposes. \cite{wu2021understanding} uses a self-encoding structure to decouple the speech signal into speaker-related privacy factors and speaker-independent content factors, and then applies a generative adversarial network (GAN) model to re-generate the privacy-preserving speech data. \cite{tran2022towards} shows that the top k positions of the generated features have higher privacy risk compared to other positions after the speech passes the privacy risk saliency estimator. Thus, Laplace noise is added to these positions to protect privacy. \cite{timm2021privacy} exploits a bottleneck layer to downscale the original features and combines it with the two-step adversarial training to remove the ASR information in the model. According to our survey, it is found that many approaches focus on ASR as the main task and remove the identity or gender information, which cannot fully satisfy the needs of SLU privacy protection algorithms.

\section{Privacy-preserving SLU}
    The proposed SLU privacy-preserving model with hidden layer separation and adversarial training, and the evaluation schemes are described in detail in this section. To reflect the hierarchy of the experiment and verify the validity of the proposed method, we present the experiments in the following two parts: 1) Separate out only the SLU task in the hidden layer, 2) Separate all three tasks in the hidden layer.
    
    \subsection{Separate Out the SLU Task Only}
	
    In this subsection, we evaluate the performance of the privacy-preserving SLU framework with separate hidden layer for SLU (SH-PPSLU). Here, the SLU task uses a specific part of the hidden layer, but the ASR as well as IR tasks still utilize the whole hidden layer. Therefore, the  SLU part of the hidden layer will still contain incomplete ASR and IR information. The desensitization capability is enhanced by incorporating adversarial training. The detailed structure of the model and training process are given in the next section (Section 4).

\begin{figure*}
	\begin{center}
		\includegraphics[width=0.95\linewidth]{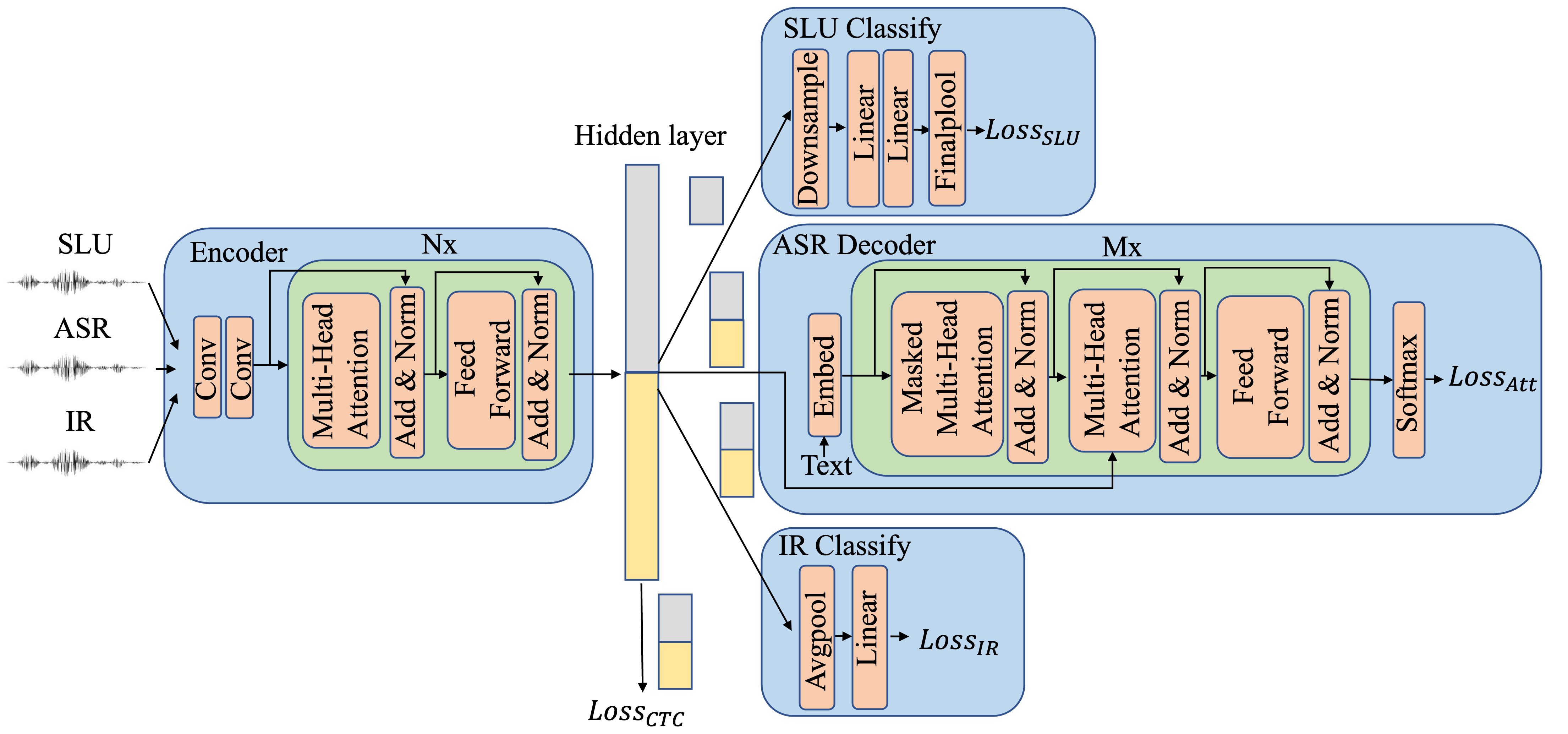}
	\end{center}
	\caption{Diagram of the SH-PPSLU model.}
	\label{fig:picture3}
    \end{figure*}

    Figure \ref{fig:picture3} plots the diagram of the proposed SH-PPSLU model. The speech signals for SLU, ASR and IR are denoted as $x_{s}$, $x_{a}$ and $x_{i}$. The joint CTC/attention transformer model \cite{deng2021improving}, commonly adopted in speech recognition, is chosen to be the encoder and ASR decoder. They will be denoted as E and AD for simplicity. In the forward propagation of the SLU training, $x_{s}$ is first passed through encoder E to compute the output of the hidden layer,$E(x_{s})$. Then, we select the first n elements of $E(x_{s})$ as the dimensionality of the SLU feature in the hidden layer, which is represented as ${E(x_{s})}_{n}$. Finally, ${E(x_{s})}_{n}$ is fed to SLU Classify, the SLU classification module to find the output vector (see the top of Figure 3). Different from \cite{lugosch2019speech}, the SLU classifier uses a structure with linear layers plus a pooling layer only. We employ the cross entropy  $L_{slu}$ as the training loss function for the SLU task.

    We pick the same amount of ASR training data as in the SLU task, in order to ensure the data balance in multi-task training. We pre-train the encoder of SH-PPSLU, E, and ASR decoder, AD, with large-scale data, which would render the ASR attacker powerful enough and also enhance the SLU performance. Unlike SLU training, the vector input to ASR decoder, AD, uses all the elements from the hidden layer. The loss function for ASR training is $L_{asr}\  =\  \alpha L_{att}+(1-\alpha )L_{CTC}$. It is a multi-task training loss function, and $\alpha$ is an adjustable parameter \cite{watanabe2017hybrid}.

    The IR model is chosen via referring to the Deepspeaker architecture \cite{li2017deep}. To fit into the framework of multi-task training, the original Deepspeaker model based on the CNN architecture is replaced using the one with the  transformer structure based on the self-attention mechanism. The IR task, like the ASR task, also uses all the features from the hidden layer for training. We shall use $L_{ir}$ to denote the Triplet loss function of IR.

    In this paper, we use (1) as the loss function for SH-PPSLU training. The weights for individual task can be adjusted according to the actual situation. It can be expected that the information of three tasks will be mixed in the hidden layer of SH-PPSLU after the multi-task training, which can be captured by ASR and IR attackers to deduce the sensitive user information. On the other hand, the hidden layer part for SLU will not only contain the SLU information, but also will include a small amount of ASR and IR information. That is, the SLU part of the hidden layer has a better privacy protection capability, compared with the whole hidden layer.
    \begin{align}
	L_{all}\  =\  {\lambda_{1} L}_{slu}  +{\lambda_{2} L}_{asr}  +{\lambda_{3} L}_{ir}
    \end{align}%

    Since the privacy-preserving capability of the SH-PPSLU model needs to be further enhanced, the adversarial training is incorporated into this model, and the privacy-preserving SLU model with hidden layer separation for SLU and adversarial training (SHA-PPSLU) is developed. The SHA-PPSLU model is obtained by fine-tuning the obtained SH-PPSLU model through applying the adversarial training loss shown in (2) below and ensuring that only the parameters of the encoder E are updated during backpropagation. This can gradually remove the ASR and IR information from the encoder E under the premise of having a robust attacker. 
    \begin{align}
		L_{adv}={{min}_{\theta_{E} }  (\lambda_{1} L}_{slu}  -{\lambda_{2} L}_{asr}  -{\lambda_{3} L}_{ir}  )
	\end{align}%
    In (2) $\theta_{E}$ represents the parameter vector of the encoder E in the SHA-PPSLU model, and $\lambda$ is an adjustable hyperparameter.

    \subsection{Complete Separation of SLU, ASR and IR}

    Note that SH-PPSLU only reserves part of the hidden layer for SLU but this part still retains certain sensitive user information. This could be mitigated through employing the adversarial training to remove significant amount of private information. In this section of experiments, we present the privacy-preserving SLU with hidden layer separation (H-PPSLU), which completely separates the three tasks in the hidden layer. With this model, the part of hidden layer for SLU can have better privacy protection capability. The training process of H-PPSLU is as follows.

    Figure \ref{fig:picture4} below shows the overall architecture of the H-PPSLU model. It is evident that H-PPSLU differs from SH-PPSLU in Figure 3 mainly in the design of the hidden layer. In particular, we can see that the hidden layer is partitioned into four parts, three of which are allocated to the three tasks in consideration. They are denoted as ${E(x_{s})}_{m}$, ${E(x_{a})}_{k}$ and ${E(x_{i})}_{l}$, where $m$,$k$,$l$ denote the size of each part. The remaining part is shared among the three tasks, SLU, ASR and IR, and it is denoted as ${E(X)}_{c}$. This shared part is introduced because it is noted that there would be some correlation between the three tasks and not including a shared part in the hidden layer would result in a significant performance degradation.

    The dimensionality of the four parts is set to 64. Of course, different setups can be adopted. In the training process, the ASR initialization is applied first, and then the features from the individual and shared parts are input to the corresponding classifier and decoder for training. Specifically SLU Classify utilizes $concat({E(x_{s})}_{m},{E(X)}_{c})$, ASR decoder AD as well as CTC loss accepts $concat({E(x_{a})}_{k},{E(X)}_{c})$, and IR Classify exploits $concat({E(x_{i})}_{l},{E(X)}_{c})$.

    \begin{figure}
	\begin{center}
		\includegraphics[width=1.03\linewidth]{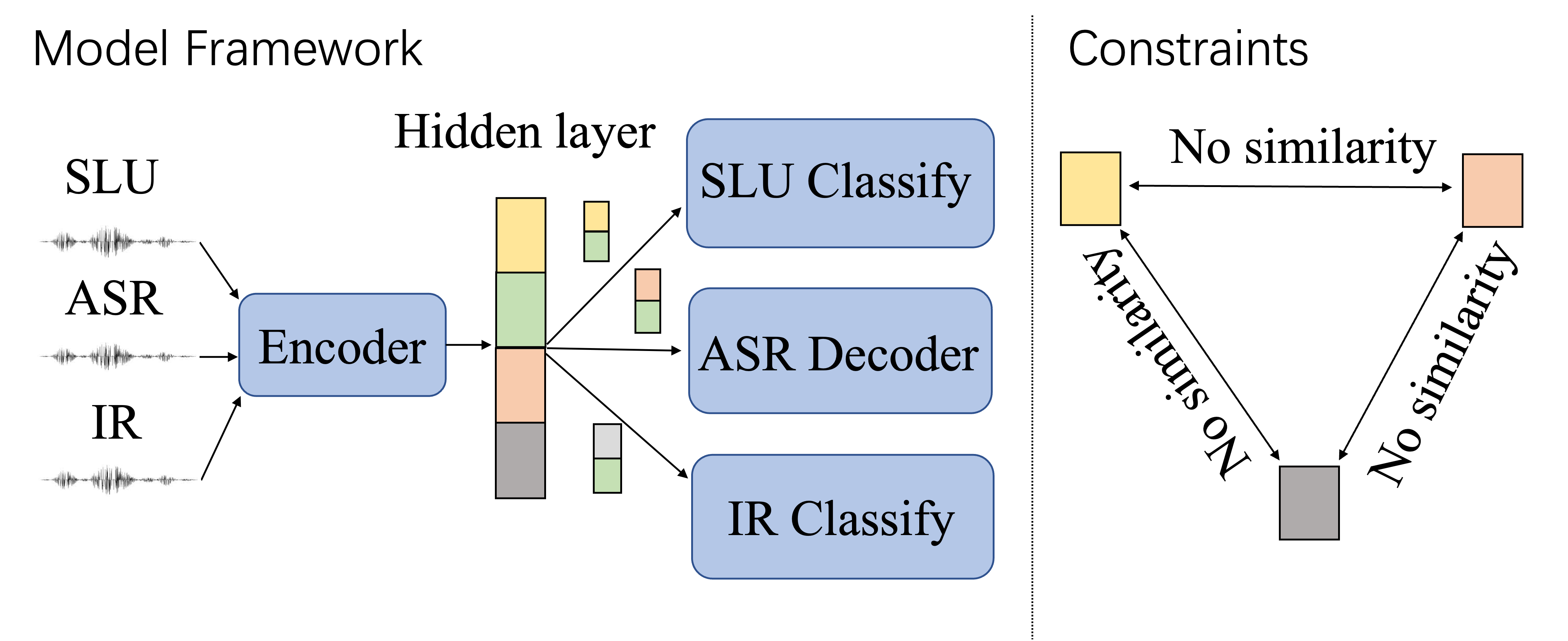}
	\end{center}
	\caption{Diagram of the H-PPSLU model.}
	\label{fig:picture4}
    \end{figure}
    We need to ensure that the correlation of the three tasks can be accounted for in the shared part of the hidden layer and the individual part contains as little information for other tasks as possible. For this purpose, we incorporate the cosine
    similarity into the loss function to ensure that the information of SLU is restrained in the individual part of the hidden layer for SLU only. The cosine similarity quantifies the similarity of two vectors A and B by calculating the cosine of their angles, as shown below in (3).

    The training loss function of the H-PPSLU model is shown in (4). In (5) ${sim}_{s,i}$ represents the cosine similarity of the vectors ${E(x_{s})}_{m}$ and ${E(x_{i})}_{l}$, and the rest of the notations are defined in a similar manner. With the above training method, SLU part of the hidden layer will have less sensitive information than that in the SH-PPSLU model, i.e., we achieve better privacy protection.
    \begin{align}
		sim=\frac{A\cdot B}{\| A\| \  \| B\| } =\frac{\sum^{n}_{i=1} {A_{i}\times B_{i}}  }{{(\sum^{n}_{i=1} {(A_{i})}^{2}  )}^{1/2}  \times {(\sum^{n}_{i=1} {(B_{i})}^{2}  )}^{1/2}  } 
    \end{align}%
    \begin{align}
		L_{all}={\lambda_{1} L}_{slu}  +{\lambda_{2} L}_{asr}  +{\lambda_{3} L}_{ir}  +{\lambda_{4} sim}_{x,y}  
    \end{align}%
    \begin{align}
    		{sim}_{x,y}  =\  {sim}_{s,i}  \  +\  {sim}_{s,a}  \  + \ {sim}_{i,a}  
    \end{align}%

    For the sake of experimental comparability and to verify whether the adversarial training can also continue to improve the privacy expectation ability of H-PPSLU, the privacy-preserving SLU with hidden layer separation and adversarial training (HA-PPSLU) is proposed. Its training process is the same as before, and the adversarial loss function is in (2).

    \subsection{Test Program Description}

\begin{figure*}
	\begin{center}
		\includegraphics[width=0.65\linewidth]{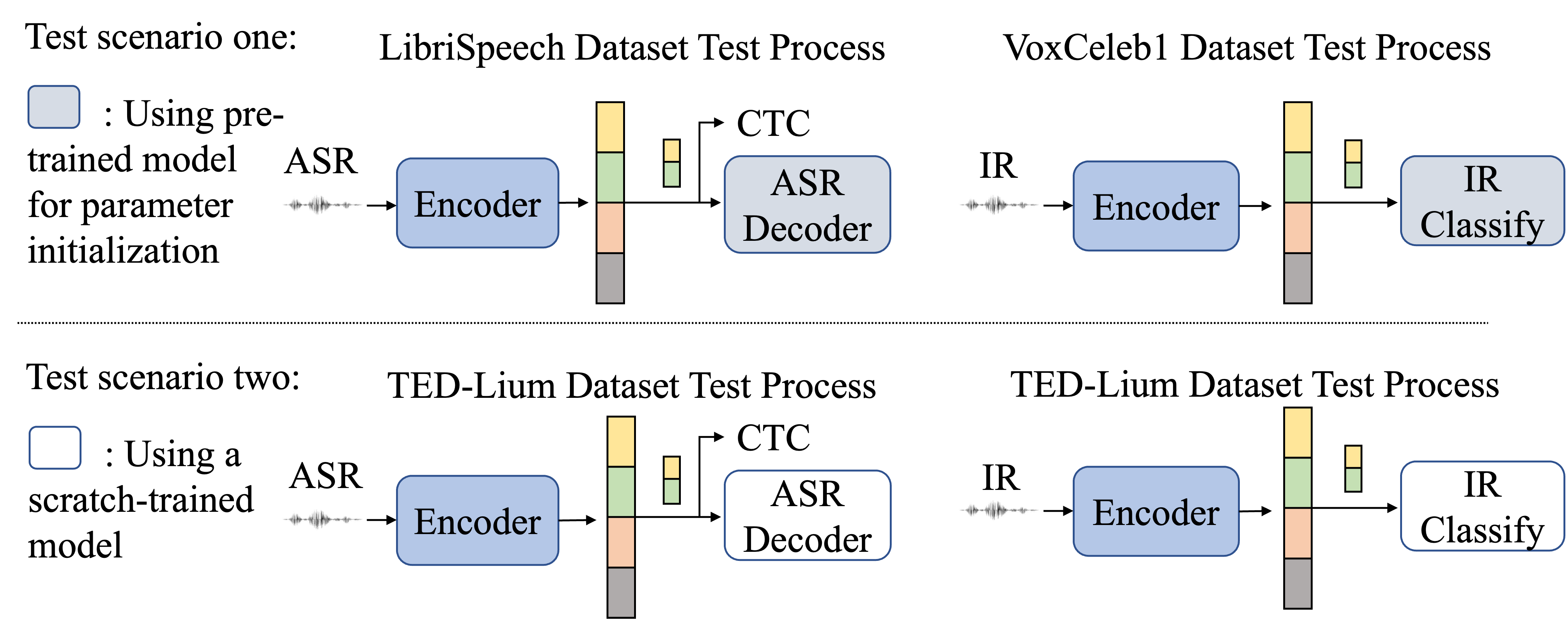}
	\end{center}
	\caption{Two testing/attacking scenarios.}
	\label{fig:picture5}
    \end{figure*}

    From the descriptions in Sections 3.1 and 3.2, it is clear that the individual part of the hidden layer for SLU guarantees the performance of SLU while containing little sensitive information. Two test scenarios will be used to verify the security of the proposed model. Figure \ref{fig:picture5} shows the specific process of the two test scenarios. In the first test scenario, to ensure a strong attacker, the encoder E, ASR decoder AD and IR Classify are tested using the corresponding trained models for parameter initialization. For testing, the test set of ASR or IR is passed through the encoder to produce the output of the hidden layer, after which the individual part for SLU in the hidden layer (i.e., the first $n$ elements in the hidden layer) is used as the input vector of the ASR Decoder AD or IR Classify, and finally the result is obtained.
    
    The second test scenario differs from the first one in that 1) the attackers of ASR as well as IR are trained from scratch after random initialization of parameters, and the parameters of the encoder E are guaranteed not to be updated during the training process. 2) Considering the actual scenario, we choose another batch of ASR as well as IR datasets for this test scenario.

\section{Experiments}
    In this section, we describe the experimental datasets, model framework setup, and experimental results. First, we expose the vulnerability of the SLU model. Then, we follow the testing scheme to evaluate whether proposed model can achieve a balance between maintaining SLU accuracy and privacy. We not only evaluate the privacy-preserving ability of the model in terms of whether the success rate of the attacker can be reduced to that of a random guess, but also compare the adversarially trained models to show that the proposed methods can further get rid of the privacy information.
    \subsection{Datasets}
    The five datasets were used to simulate realistic scenarios, namely SLURP \cite{bastianelli2020slurp}, Fluent Speech Commands (FSC) \cite{lugosch2019speech}, LibriSpeech \cite{panayotov2015librispeech},Voxceleb1 \cite{nagrani2017voxceleb} and TED-Lium \cite{rousseau2012ted}.

    \paragraph{LibriSpeech.} LibriSpeech is a large dataset consisting of approximately 1000 hours of English reading. It is derived from the reading of audiobooks from the LibriVox project. In this paper, we use train-clean360 and train-clean100 to pre-training and training our ASR task.
    
    \paragraph{VoxCeleb1.} The VoxCeleb1 dataset contains more than 100,000 discourses from 1,251 celebrities, extracted from videos uploaded to YouTube. These speakers have different accents, occupations and ages.

    \paragraph{FSC.} The FSC dataset consists of a single channel 16KHZ wav audio file, which includes approximately 19 hours of voice duration. Each audio file contains records of English voice command words that may be used in a smart home or virtual assistant. The dataset contains a total of 31 intents.

    \paragraph{SLURP.} SLURP is a publicly available multi-domain E2E-SLU dataset. It is much larger and more diverse than existing SLU datasets. Its collection of about 72,000 audio recordings includes more than 18 different scenes with 46 defined actions and 55 different entity types.

    \paragraph{TED-LiUM.} TED-LIUM is a training corpus for automatic speech recognition from TED talks, which containing in total about 118 hours of talks. Not only we use this dataset for ASR testing, also adapt the dataset for IR testing, which contains 774 speakers and 13,296 data pairs.

    \subsection{Setup and Evaluation Metrics}
    Our experiments were all conducted on the open source tool, i.e., Espnet \cite{watanabe2018espnet}. First, the 80-dimensional Fbank features are extracted as low-dimensional feature inputs, and its frame length is 25ms with 10ms frame shift. In order to make the input size of the model match the selected data volume, We set the number of layers for Encoder and Decoder to be 12 and 6 respectively, and the attention dimension, number of header as well as droupout ratio to be 256, 4 and 0.1. Then, the individual part dimension of the SLU of the SH-PPSLU model is set to be 128, and the individual and shared part dimensions of the three tasks of the H-PPSLU model are both 64. 
    Our model is initialized with a 50-epoch trained ASR model weight. During the training process, the algorithms are trained for 15 epochs using Adam optimizer with the learning rate 0.001, and the coefficient of the ASR loss function 0.1. For the adversarial training we train the model for 10 epochs.

    For the evaluation of SLU and IR, the classification accuracy is calculated in this paper and denoted as ACC-SLU and ACC-IR, respectively. For the ASR evaluation, we compute the word error rate (WER) for a given test set. The higher the value, the better the privacy-preserving ability for spoken contents.

    \subsection{Experimental Results of Test Scenario I}
    In Table \ref{table1}, ML-SAI is the base model without privacy protection for the SLU, ASR, and IR multi-task training. SH-PPSLU, SHA-PPSLU, H-PPSLU-nocos, H-PPSLU, and HA-PPSLU are the five privacy-preserving SLU models proposed in this paper. H-PPSLU-nocos is the H-PPSLU model without the cosine similarity constraint during the training process. AT-SAI,AT-SI, PP-WWV, VAE are four comparison algorithms. Note that the last three methods only protect one type of the private information.

    From the experimental results, we can see that our proposed method of using only a portion of the features of the hidden layer to train the back part of the model does not cause a significant degradation in the performance of the SLU task, which is basically maintained at the same level as that of the baseline. It shows that the SLU classifier performs well in the test datasets only using a specific dimension feature in the hidden layer. When the model training is combined with adversarial training, it causes a small drop (two points) in the SLU performance, which is acceptable. We can also find that the SLU performance degradation is smaller using the FSC dataset than that using the SLURP dataset. Overall, our proposed model can maintain the performance of SLU well.
    \renewcommand{\tabcolsep}{2.0pt}
    \begin{table*}[t]
        \footnotesize
    	\begin{center}
    			{
    				\renewcommand{\arraystretch}{1.10}
    			\begin{tabular}{c|ccc|ccc}
    				\hline
    				&&SLURP &&&FSC    \\ \hline
    				Method	&ACC-SLU$\uparrow$ &WER-ASR$\uparrow$ &ACC-IR$\downarrow$ &ACC-SLU$\uparrow$ &WER-ASR$\uparrow$ &ACC-IR$\downarrow$ \\ \hline
                    ML-SAI (w/o privacy protection)	&74.1	&12.6	&82.8	&99.4	&12.0	&83.1 \\\hline
                    AT-SAI	&72.8	&69.1	&54.3	&98.2	&77.1	&55.0 \\ 
                    SH-PPSLU	&73.8	&49.7	&69.8	&99.3	&65.3	&77.2 \\
                    SHA-PPSLU	&72.1	&78.6	&53.5	&98.0	&88.0	&53.0 \\ 
                    H-PPSLU-nocos	&\bf 73.9	&75.3	&69.0	&\bf99.2	&78.5	&69.5 \\
                    H-PPSLU	&73.4	&87.4	&66.2	&99.1	&86.8	&67.2 \\
                    HA-PPSLU	&72.2	&\bf89.8	&\bf52.2	&97.2	&\bf92.1	&\bf52.3 \\ 
                    AT-SI \cite{srivastava2019privacy}	&72.9	&-	&53.8	&98.2	&-	&54.0 \\
                    PP-WWV \cite{timm2021privacy}	&71.4	&81.9	&-	&97.1	&85.4	&- \\
                    VAE \cite{stoidis2021protecting}	&72.7	&-	&54.4	&98.3	&-	&55.7 \\ \hline              
    			\end{tabular}
    		}
    		\caption{Comparisons of performance among algorithms using different datasets with and without privacy protection. We use metrics (WER-ASR and ACC-IR) to measure privacy protection capabilities. For our goal in this paper, $\uparrow$ represents the higher the value, the better the performance, $\downarrow$ denotes the opposite of the former.}\label{table1}
    	\end{center}
    \end{table*}

    \renewcommand{\tabcolsep}{2.0pt}
    \begin{table*}[t]
        \footnotesize
    	\begin{center}
    			{
    				\renewcommand{\arraystretch}{1.10}
    			\begin{tabular}{c|ccc|ccc}
    				\hline
    				&&SLURP &&&FSC    \\ \hline
    				Method	&ACC-SLU$\uparrow$ &WER-ASR$\uparrow$ &ACC-IR$\downarrow$ &ACC-SLU$\uparrow$ &WER-ASR$\uparrow$ &ACC-IR$\downarrow$ \\ \hline
                    ML-SAI (w/o privacy protection)	&74.1	&12.6	&82.8	&99.4	&12.0	&83.1 \\\hline 
                    H-PPSLU($c$=88)	&\bf73.8	&78.5	&67.6	&\bf99.2	&80.2	&67.3 \\ 
                    H-PPSLU($c$=76)	&\bf73.8	&84.0	&67.1	&99.0	&85.9	&67.3 \\
                    H-PPSLU($c$=64)	&73.4	&87.4	&66.2	&99.1	&86.8	&67.2 \\ 
                    H-PPSLU($c$=52)	&73.2	&\bf89.0	&65.3	&99.0	&\bf90.1	&66.8 \\
                    H-PPSLU($c$=40)	&72.8	&86.1	&\bf63.7	&99.1	&86.5	&\bf62.6  \\ \hline           
    			\end{tabular}
    		}
    		\caption{Comparisons of the SLU, IR and ASR accuracy of H-PPSLU for different shared dimensions on different datasets. $c$ denotes the dimension of the shared part.}\label{table2}
    	\end{center}
    \end{table*}

    As seen in Table 1, the SLU model is a vulnerable deep learning model with a potential hidden layer representation that can be easily stolen by attackers and used to reason about the users' sensitive information. We use metrics (WER-ASR and ACC-IR) to measure privacy protection capability. 
    The higher the value of WER-ASR, the better the privacy protection ability for the speech content. The smaller the value of ACC-IR, the better the voiceprint privacy protection ability (for the 1:1 IR verification scenario, 50\% is the lowest performance value).
    
    From Table 1, we can see that our proposed privacy-preserving scheme has better privacy-preserving ability than others. From experiment results, the ASR task relies more on all features of the hidden layer, while IR can achieve better accuracy using only a portion of the features of the hidden layer. The results from the 6$th$ row show that training the model combining with adversarial training further improves the privacy protection capability. From the results of H-PPSLU-nocos and SH-PPSLU, it can be seen that the specific parts of the hidden layer for SLU, ASR, and IR respectively help improve the privacy protection, especially for ASR. From the results in the 7$th$-9$th$ rows, we can see that adding cosine dissimilarity helps to reduce the attacker's recognition accuracy. 
    The H-PPSLU model can get rid of most ASR information, but there is still some room for improvement in protecting the IR information. Finally, it can be seen that the model can better reduce the success rate of IR attackers when adversarial training is added in the training process. 
    
    Overall our proposed privacy-preserving model can well reduce the inference success rate of malicious attackers to that of a random guess while ensuring that the SLU performance is not affected (we consider when WER is approximately close to 100\% and ACC-IR close to 50\% the outputs of the ASR and IR task are approximately close to random guesses). Combining with adversarial training is effective in improving the privacy preservation ability and achieving consistent convergence using both of the SLU datasets.

    Since the shared dimensions in the previous experiments were all 64, in order to verify the effect of different shared dimensions on the privacy-preserving ability of the H-PPSLU model. We adopt different values of the shared dimensions $c$ (i.e., 88, 76, 64, 52 and 40, respectively) to carry out the corresponding experiments. For above five experiments, the individual parts of the three tasks have the same dimensionality. 
    
    From Table \ref{table2}, we can see that the smaller $c$ is, the lower the accuracy of SLU is, but the accuracy id changed within one percentage point. 
    With the value change of $c$, WER-ASR and ACC-IR also have relatively small changes.From these observations, we find that the change of the shared dimension $c$ has little influence on the effect of privacy protection. 
     \subsection{Experimental Results of Test Scenario II}
    \renewcommand{\tabcolsep}{2.0pt}
    \begin{table}[t]
        \footnotesize
    	\begin{center}
    			{
    				\renewcommand{\arraystretch}{1.10}
    			\begin{tabular}{c|cc|cc}
    				\hline
    				&SLURP &&FSC    \\ \hline
    				Method &WER-ASR$\uparrow$ &ACC-IR$\downarrow$  &WER-ASR$\uparrow$ &ACC-IR$\downarrow$ \\ \hline
                    ML-SAI	&22.1	&90.5	&23.3	&90.2 \\\hline
                    AT-SAI	&82.6	&71.8	&83.0	&77.8 \\ 
                    H-PPSLU	&42.5	&75.3	&39.2	&77.5 \\
                    HA-PPSLU	&\bf92.1	&\bf60.2	&\bf92.3	&\bf68.5 \\ \hline         
    			\end{tabular}
    		}
    		\caption{Attack experiment results under the test scenario 2.}\label{table3}
    	\end{center}
    \end{table}

    In the actual attack scenario, the attacker can use more powerful attack means, that is, training the attack model from scratch, as shown in Scenario 2 in Figure 5. In addition, attackers generally do not know the data set used by our training model. For this reason, we use different training data (ASR: TED-LIUM, IR: TED-LIUM-I) to simulate the behavior of attackers.
    We use the encoder model in the first test scenario for the test of the scenario 2. The IR classifier as well as ASR decoder are trained with random initialization. The test results can be seen in Table \ref{table3}. Under this attack, the attacker is able to obtain relatively high IR errors, but the accuracy of ASR is still maintained at a low level. We also find that the adversarial training further enhances the privacy-preserving ability of the hidden layer separation scheme. In other words, the hidden layer separation method also improves that of the adversarial learning. Overall, our model still has the better privacy protection capability under this attack.
    
\section{Conclusion}
    We propose a novel hybrid privacy-preserving algorithm that uses the hidden-layer feature separation and adversarial training framework for spoken language understanding. First, we divide the hidden layer into several parts according to the number of tasks, each of which occupies a shared and an individual part, and constrain the correlation between tasks by the cosine similarity, so that the individual information of different tasks is distributed in different parts of the hidden layer. Then, the adversarial training is utilized to further enhance the ability of privacy protection. It is shown that our method not only maintains the SLU accuracy without significantly degradation, but also reduces the success rate of attackers close to that of a random guess.

\bibliographystyle{named}
\bibliography{ijcai23}
\end{document}